\documentclass[twocolumn]{revtex4-2}
\usepackage{graphicx}
\usepackage{dcolumn}
\usepackage{bm}
\usepackage{amsmath}
\usepackage{amsfonts}
\usepackage{amssymb}
\usepackage{color}
\usepackage{epstopdf}
\usepackage{float}

\usepackage{caption}

\providecommand{\U}[1]{\protect\rule{.1in}{.1in}}
\newcommand{\ket}[1]{|#1\rangle}
\newcommand{\bra}[1]{\langle#1|}

\newcommand{\modu}[1]{\vert\vec{#1}\vert}

\begin{document}
\title{Qubit thermodynamics far from equilibrium: 
	two perspectives about the nature of heat and work 
	in the quantum regime}
\author{Andr\'es Vallejo}
\author{Alejandro Romanelli}
\author{Ra\'ul Donangelo}
\affiliation{\begin{small} Facultad de Ingenier\'{\i}a, Universidad 
de la Rep\'ublica, Montevideo, Uruguay\end{small}}
\date{\today}
\begin{abstract}
\noindent 
Considering an entropy-based division of energy transferred 
into heat and work, we develop an alternative theoretical 
framework for the thermodynamic analysis of two-level systems. 
When comparing these results with those obtained under the 
standard definitions of these quantities, we observe the 
appearance of a new term of work, which represents the energy 
cost of rotating the Bloch vector in presence of the external 
field that defines the local Hamiltonian. 
Additionally, we obtain explicit expressions for the temperature, the heat capacity and the internal entropy production of the system in both paradigms. 
In order to illustrate our findings we study, from both 
perspectives, matter-radiation interaction processes for two different systems.
\end{abstract}

\maketitle

\section{\label{sec:level1}Introduction}
Quantum physics is an intrinsically dynamic theory, and 
therefore time-dependence is essential in its description.
Classical thermodynamics, on the other hand, mostly considers
closed systems that evolve quasi-adiabatically \cite{Carnot}. 
For this reason, when we insert quantum dynamics into 
thermodynamics we obtain a quantum version of finite-time thermodynamics, which is intimately related with the theory 
of open systems\cite{Kosloff}. 

Heat and work are the basic mechanisms of energy exchange 
between thermodynamic systems. 
From the classical point of view, heat is usually defined as 
the energy flow which occurs exclusively due to the 
temperature difference between the systems. 
Work, on the other hand, is the energy exchange which can be 
measured  through the variation of a macroscopic parameter, 
such as the volume of the system \cite{Carnot,Sonntag,Zemansky,Callen,Cengel}. 

Although for classical thermodynamic systems the classification 
of the energy transfers as heat and work is not free from
controversy, the situation is even more complex when quantum 
systems are considered. 
Many non-equivalent definitions of these quantities can be found 
in the literature \cite{Bender,Jarzynski,Uzdin,Bera,Horodecki,
Skrzypczyk,Alicki}, so the correct identification of heat and work in that regime 
can be considered an open problem. 

One of the most extensively considered paradigms regarding these 
quantities was proposed by Alicki several decades ago 
\cite{Alicki}. 
Defining the internal energy of the system as the expected value 
of the local Hamiltonian $H$ in the actual reduced state $\rho$,
\begin{equation}\label{E1}
E=\langle H\rangle =\text{tr}[H\rho],
\end{equation}
thus, an infinitesimal energy change takes the form
\begin{equation}\label{dE1}
dE=\text{tr}[dH\rho]+\text{tr}[Hd\rho].
\end{equation}
\noindent The first term on the right-hand side of 
Eq. (\ref{dE1}) is the energy change due to changes in the 
Hamiltonian of the system, associated to some control parameter 
which can be modified by the experimenter. 
Considering the previous discussion about the classical work 
made on the system, it is reasonable to define the infinitesimal of work as
\begin{equation}\label{work1}
\delta \mathcal{W}=\text{tr}[dH\rho].
\end{equation}
\noindent Thus, in order to assure the validity of the first law, 
the infinitesimal of heat is defined as
\begin{equation}\label{heat1}
\delta \mathcal{Q}= \text{tr}[Hd\rho].
\end{equation}
Therefore, from this point of view, heat is related to changes 
in the density matrix describing the quantum state.

Despite its wide application in several contexts \cite{Deffner,Alicki2},
this approach has some weak points. 
For example, if we consider two interacting systems with 
constant local Hamiltonians, the above equations implies 
that, even if work is done on the global system through a 
time-dependent Hamiltonian, no work is done on the individual systems, which is counterintuitive. 

In this paper we will explore some consequences of a recent 
proposal that claims to resolve these issues \cite{Ahmadi,Alipour1}. 
It is based on the hypothesis that the von Neumann entropy 
is a valid extension of the thermodynamic entropy in the 
quantum regime, and on the fact that it depends only on the 
eigenvalues $\lambda_{j}$ of the density matrix $\rho_{S}$,
\cite{Nielsen}:
\begin{equation}\label{Svn}
S_{vN}=-\sum_{j=1}^{N}\lambda_{j}\text{ln}{\lambda_{j}}.
\end{equation} 
This implies that the changes in $S_{vN}$ are always accompanied 
by changes in the eigenvalues of $\rho_{S}$. 
But classically, the entropy change is proportional to the 
reversible heat transfer. 
These observations suggest that heat could not be related with 
changes in the whole density matrix (see Eq. (\ref{heat1})), 
but only on its eigenvalues. 

The remainder of this paper is organized as follows. 
In Section II we introduce the new notions of work and heat for a generic finite-dimensional quantum system.
In Section III, using the above ideas, we develop the complete 
thermodynamics for a two-level system and we compare the 
results with those which arise from assuming Alicki's 
theoretical framework \cite{Vallejo1}.
As we shall see, the simplicity of the two-level system allows 
a very simple geometric interpretation of the thermodynamic 
quantities.
In Section IV, we perform both thermodynamic analyses for two
specific matter-radiation interaction processes. 
Some conclusions are presented in Section V.

\section{The new paradigm}
In this section we briefly explain the ideas behind the 
new proposal. 
First, we note that, in the standard paradigm, 
Eqs. (\ref{work1}) and (\ref{heat1}) are equivalent to,
\begin{equation}\label{workP1g}
\delta \mathcal{W}=\sum_{j}\rho_{jj}dE_{j}
\end{equation}
and
\begin{equation}\label{heatP1g}
\delta \mathcal{Q}=\sum_{j}d\rho_{jj}E_{j}.
\end{equation}
where $\lbrace E_{j}\rbrace$ are the eigenenergies of the system, 
and $\lbrace \rho_{jj}\rbrace$ their corresponding probabilities. 
These expressions are equivalent to those corresponding to the 
classical notions of work and heat for systems with a discrete 
spectrum. 
But it would be reasonable that quantum features, such as the 
existence of coherence between the different eigenstates, play 
an important role, which is not covered by the previous 
description \cite{Su,Francica}.  

To analyze the problem from a different perspective, 
we start by writing the instantaneous spectral decomposition 
of the density matrix,

\begin{equation}
\label{specdec}
\rho =\sum_{j}\lambda_{j}\ket{\psi_{j}}\bra{\psi_{j}},
\end{equation}
where $\{\ket{\psi_{j}}\}$ are the eigenfunctions and  
$\{\lambda_{j}\}$ the set of corresponding eigenvalues. 
The above equation, together with Eq. (\ref{E1}), allows to 
express the internal energy as,
\begin{equation}\label{EN}
E=\sum_{j}\lambda_{j}\bra{\psi_{j}}H\ket{\psi_{j}}
\end{equation}
so the infinitesimal energy change is given by
\begin{equation}
dE=\sum_{j}d\lambda_{j}\bra{\psi_{j}}H\ket{\psi_{j}}
+\sum_{j}\lambda_{j}d\bra{\psi_{j}}H\ket{\psi_{j}}.
\end{equation}
Recalling Eq. (\ref{Svn}), it is clear that only the first term 
on the right-hand side of the above equation is linked to the 
entropy change. 
Thus, it is the only term which should be considered as heat, 
so we define,
\begin{equation}\label{heatP2g}
\delta Q=\sum_{j}d\lambda_{j}\bra{\psi_{j}}H\ket{\psi_{j}},
\end{equation} 
and, as a consequence,
\begin{equation}\label{workP2g}
\delta W=\sum_{j}\lambda_{j}d\bra{\psi_{j}}H\ket{\psi_{j}}. 
\end{equation}

Note that in this new paradigm, work is related not only 
to the possibility of driving the Hamiltonian, but also 
with the change in the eigenvectors of the density matrix. 
Of course, for thermal equilibrium states, and, more in
general, for any incoherent state in the energy basis, 
the Hamiltonian and the density matrix commute.
Thus,  $\lambda_{j}=\rho_{jj}$,  
$\bra{\psi_{j}}H\ket{\psi_{j}}=E_{j}$, and, as a consequence,
both paradigms are equivalent in that limit.

In the next section we focus on the study of two-level systems, 
a case which is interesting in its own right due to its 
technological applications, and which, due to its simplicity, 
allows to obtain a clear geometrical interpretation of the thermodynamic quantities.

\section{Thermodynamic quantities for two-level systems 
	in the Bloch vector representation}
\subsection{Internal energy, heat and work}
A convenient way to visualize the state of a two-level system 
is through its Bloch vector,
\begin{equation}
\vec{B}=(B_{x},B_{y},B_{z}),
\end{equation} 
\noindent whose components are, aside from a factor $\hbar/2$, 
the expected values of the spin operators $S_{x}, S_{y}$ and $S_{z}$,
\begin{equation}
\begin{split}
&B_{x}=\langle S_{x}\rangle = tr(\rho_{_{S}}\sigma_{x}),\\
&B_{y}=\langle S_{y}\rangle = tr(\rho_{_{S}}\sigma_{y}),\\
&B_{z}=\langle S_{z}\rangle = tr(\rho_{_{S}}\sigma_{z}),
\end{split}
\end{equation}
\noindent where $\sigma_{x}$, $\sigma_{y}$ and $\sigma_{z}$ 
are the Pauli matrices. 
On the other hand, aside from an irrelevant scalar multiple
of the identity, a generic Hamiltonian in two dimensions 
adopts the form,
\begin{equation}\label{Hamiltonian}
H=-\vec{v}.\vec{\sigma},
\end{equation}
where $\vec{v}$ can be associated to an effective magnetic 
field, and $\vec{\sigma}$ is a formal vector whose components 
are the Pauli matrices.
In terms of the Bloch vector, the density matrix of a two-level 
system can be written as,
\begin{equation}\label{rdm}
\rho=\frac{1}{2}[1+\vec{B}.\vec{\sigma}].
\end{equation}
\noindent Using Eqs. (\ref{E1}), (\ref{Hamiltonian}), (\ref{rdm})
and the identity
\begin{equation}\label{identity}
(\vec{a}.\vec{\sigma})(\vec{b}.\vec{\sigma})
=(\vec{a}.\vec{b})I+i\vec{\sigma}.(\vec{a}\times\vec{b}),
\end{equation}
\noindent we obtain the internal energy
\begin{equation}\label{internal energy}
E=-\vec{B}.\vec{v}.
\end{equation}

From the above equation we can write an infinitesimal energy 
change as:
\begin{equation}
dE=-d\vec{B}.\vec{v}-\vec{B}.d\vec{v}.
\end{equation}

\subsubsection{Standard framework: Paradigm 1}

In the standard framework, work is performed on the system 
only if the Hamiltonian is time-dependent. 
Heat, on the other hand, is related to changes in the quantum 
state. 
This point of view leads to the following natural definitions 
of infinitesimal work and heat for a two-level system,
\begin{equation}\label{workP1}
\delta\mathcal{W}=-\vec{B}.d\vec{v},
\end{equation}  
\begin{equation}\label{heatP1}
\delta\mathcal{Q}=-d\vec{B}.\vec{v},
\end{equation}
\noindent in such a way that the first law has the same 
structure than in the classical case: 
$dE=\delta\mathcal{Q}+\delta\mathcal{W}.$ 

From this point of view, the work is zero when the effective 
magnetic field is constant in time, or when its change is 
orthogonal to the Bloch vector, i.e, orthogonal to the 
instantaneous magnetization. 
Conversely, heat is zero when the Bloch vector is constant in time, i.e. the system is in equilibrium, and also when its 
change is orthogonal to the effective magnetic field. 
This situation includes the special case in which the qubit 
evolves unitarily.
 
\subsubsection{Alternative treatment: Paradigm 2}

To analyze the problem from this new perspective, we first 
note that the eigenvalues of $\rho_{S}$ can be written as
\begin{equation}\label{eigenvalues}
\lambda_{+/-}=1/2\pm B/2,
\end{equation}
\noindent where $B$ is the modulus of the Bloch vector.
Therefore, considering  Eq. (\ref{Svn}), we can write the 
von Neumann entropy of the qubit as,
\begin{small}
\begin{equation}\label{entropy1}
\frac{S_{vN}}{k_{B}}=
-\left(\frac{1+B}{2}\right)\ln\left(\frac{1+B}{2}\right)
-\left(\frac{1-B}{2}\right)\ln\left(\frac{1-B}{2}\right).
\end{equation}
\end{small}
\noindent We note that in the case of two-level systems, 
the entropy depends only on $B$, so its changes are linked 
solely to changes in $B$. 
Thus, from the point of view according to which heat 
is associated to the energy exchange responsible for  
entropy change, we conclude that $\delta Q$ is different 
from zero if and only if $B$ changes. 

If we write the energy of the system, 
Eq. (\ref{internal energy}), as
\begin{equation}\label{EP2}
E=-B\hat{B}.\vec{v},
\end{equation}
where $\hat{B}$ is a unit vector in the direction of $\vec{B}$, 
the energy change can be partitioned as
\begin{equation}
dE=-dB\left(\hat{B}.\vec{v}\right)-Bd\left(\hat{B}.\vec{v}\right).
\end{equation}
\noindent As a result of the previous considerations, and
since only the first term contributes to the entropy change, 
we define the heat and work exchanged in this second 
approach as
\begin{equation}\label{heatP2}
\delta Q \equiv -dB\left(\hat{B}.\vec{v}\right),
\end{equation}
\noindent and
\begin{equation}\label{workP2}
\delta W \equiv -Bd\left(\hat{B}.\vec{v}\right).
\end{equation}

It is interesting to consider in detail these expressions.
From Eq. (\ref{workP2}), we notice that the work done 
on the system is the product of the modulus of the magnetization 
$B$ and the infinitesimal change in the projection of the 
effective magnetic field on the direction of the magnetization,
\begin{equation}
\delta W=-Bd\left(\vert \vec {v}\vert\cos \theta \right),
\end{equation}
\noindent where $\theta$ is the angle between $\hat{v}$ 
and $\hat{B}$. 
Thus, in this framework, work can be performed on the 
system even if the Hamiltonian, i.e., $\vec{v}$, is fixed, 
provided that $\hat{B}$ changes in such a way that the 
angle between both vectors is not constant in time. 
In particular, work is extracted from the system, 
$\delta W\leq 0$, when the Bloch vector changes in such a way 
that $\hat{B}.\vec{v}$ increases in time, i.e. when $\vec{B}$ 
tends to align itself with the effective field $\vec{v}$.

On the other hand, there are two kind of processes for which 
no heat is exchanged: 1) isoentropic processes, for which $B$ 
is constant, and 2) processes for which $\hat{B}\perp\vec{v}$ along the process. 
This situation includes states that encompass all possible entropy values, as long as the point representing the reduced 
state in the Bloch sphere moves on a plane orthogonal to the 
effective magnetic field. 
The states located on this plane are the statistical mixtures 
of the cat-states (SMCS) of the instantaneous Hamiltonian, 
and all of them possess zero energy, so if the system transits 
among these states the work exchanged is also zero. 

Regarding the sign of the heat, it is clear that if the 
angle $\theta$ between $\hat{B}$ and $\vec{v}$ satisfies 
$\theta < \pi/2$, the injection of heat into the system, 
$\delta Q >0$, implies an increase in entropy. 
Conversely, if $\theta > \pi/2$, injection of heat leads to an 
entropy decrease. 
This suggests that temperatures associated to states in 
the upper hemisphere of the Bloch sphere take positive values, while those on the other hemisphere have opposite sign.
This point which will be studied in detail in the next subsection.
\subsubsection{Discussion}
The relation between heat and work in the first paradigm, 
$\mathcal{Q}$ and $\mathcal{W}$, and in the second, $Q$ and $W$,
can be obtained as follows. 
From Eq. (\ref{workP2}),
\begin{equation}
\delta W=-Bd\left(\hat{B}.\vec{v}\right)
=-Bd\hat{B}.\vec{v}-B\hat{B}.d\vec{v},
\end{equation}
\noindent and since 
$-B\hat{B}.d\vec{v}=-\vec{B}.d\vec{v}=\delta\mathcal{W}$, 
we obtain:
\begin{equation}\label{works}
\delta W=\delta\mathcal{W}-Bd\hat{B}.\vec{v}.
\end{equation}
\noindent Similarly, the relation between the heats exchanged 
in  both paradigms is:
\begin{equation}\label{heats}
\delta Q=\delta\mathcal{Q}+Bd\hat{B}.\vec{v}.
\end{equation}  
Eq. (\ref{works}) can be written as
\begin{equation}\label{totalW}
\delta W=\delta\mathcal{W}+\delta W' ,
\end{equation}
where
\begin{equation}\label{deltaW}
\delta W'=-Bd\hat{B}.\vec{v}.
\end{equation}  

We notice that the work associated to an infinitesimal 
process in paradigm 2 adds, to the standard contribution $\delta\mathcal{W}$ due to Hamiltonian driving, the additional 
term $\delta W'$, which is related to the time variation of 
the density matrix eigenvectors in Eq. (\ref{workP2g}). 
Choosing the $z$ axis in the direction of $\vec{v}$, and 
expressing $d\hat{B}$ in spherical coordinates,
\begin{equation}
d\hat{B}=d\theta \hat{e}_{\theta}+\sin\theta d\varphi\hat{e}_{\varphi},
\end{equation}
\noindent we find that
\begin{equation}
\delta W'=B\sin\theta\varepsilon d\theta,
\end{equation}
\noindent where $\varepsilon =\modu{v}$ is the positive 
energy eigenvalue. 

Since the components of the Bloch vector are proportional to the 
expected values of the spin operators, $\vec{B}$ can be 
interpreted as the average magnetic dipole moment of the system, 
which is embedded in an external magnetic field $\vec{v}$. 
For a classical dipole, the potential energy is given by 
$U=-\vec{B}.\vec{v}$, so it coincides with our expression 
for the internal energy, Eq. (\ref{internal energy}).
Therefore, the work that must be performed against the magnetic 
field in order to rotate the dipole from the initial to the final configuration equals the potential energy difference  between 
those two configurations. 

In our case we notice that, unlike what occurs in unitary evolutions, the interaction with the environment drives the 
system along trajectories such that the polar angle $\theta$ 
may vary in time, which implies that rotational work must be 
performed against the magnetic field, in an amount:
\begin{equation}
\delta W_{rot}=-\vec{M}.\vec{d\theta}
\end{equation} 
where $\vec{M}=\vec{B}\wedge\vec{v}$ is the torque exerted by the magnetic force, and $\vec{d\theta}=d\theta\hat{e_{\varphi}}$.
It is now straightforward to show that 
$\delta W'= \delta W_{rot}$.
Thus, $\delta W'$ is the energetic cost of rotating the dipole 
in the presence of the external field. 

We also notice that $B\sin\theta$ is the coherence of the state 
measured using the $l_{1}$ norm, $C_{l_{1}}$ \cite{Baumgratz}, 
\begin{equation}
C_{l_{1}}\equiv\sum_{i\neq j}\vert \rho_{ij} \vert =B\sin\theta.
\end{equation} 
\noindent From the above equations, we obtain
\begin{equation}\label{workcoherence}
\delta W'=C_{l_{1}}\varepsilon d\theta,
\end{equation}
\noindent so we conclude that for a fixed local Hamiltonian, 
work can be performed on the system only if coherence 
in the energy eigenbasis is present. 
In fact, from Eq. (\ref{workcoherence}) we see that $C_{l_{1}}$  
can be interpreted as the lever arm of the torque, revealing, 
in the context of the present work, the role of 
quantum coherence as a resource for thermodynamic tasks 
\cite{Scully,Korzekwa}. 
Reciprocally, in Ref. \cite{Abiuso} it is shown, employing 
a differential geometry approach, that the creation of 
coherence is detrimental to efficiency in finite-time 
thermodynamic processes.

As a simple example let us consider a qubit undergoing a pure 
dephasing process. 
Since the Hamiltonian is fixed, from the point of view of 
paradigm 1 no work is performed. 
Since the non-diagonal terms tend to zero while the populations 
remain constant, the Bloch vector evolves in such a way that the 
energy of the system does not change. 
This implies that no heat is exchanged either, so from the 
point of view of paradigm 1, pure dephasing is a 
non-dissipative process in which the information contained 
in the coherence is transferred from the system to the environment. 

However, information possesses an energy value 
\cite{Szilard,Toyabe,Peterson}, so it should be expected that, 
despite maintaining its energy constant, the potential of the qubit to do work would decrease. 
This fact can be explained in a natural way analyzing the problem 
from the point of view of paradigm 2. 
Since during pure dephasing the entropy of the qubit increases, 
for positive temperature states, heat is transferred to the 
system. 
But since the energy of the system is constant, an equal but opposite amount of work is performed on the environment. 
This decrease in the ability to perform work can then be 
interpreted, as the result of giving some high quality energy 
(work), receiving, in exchange, the same amount of low quality energy (heat). 

\subsection{Temperature}

Our main objective so far has been to compare the notions
of heat and work within each of the two paradigms considered, 
However, the adoption of either one allows us to extend,
in the case of two-level systems, other thermodynamic 
quantities to the quantum regime, considering their 
corresponding classical analogous concepts.

Temperature is clearly defined only for 
systems in thermodynamic equilibrium. 
Nevertheless, many definitions of temperature have been 
shown to be useful in non-equilibrium situations \cite{Gemmer,Johal,Poliblanc,Williams, Latune,Brunner2,Manatuly,Roman,Ali}.

The temperatures that we define below should be interpreted 
as a measure of the entropy changes produced by the heat 
exchanged when the system finds itself in a particular state.
They are not necessarily linked to the direction of the heat 
flow when thermodynamic systems are put in thermal contact. 
In fact, it has been theoretically predicted and experimentally 
shown that the direction of the heat flow between quantum 
systems in local thermal states can be reversed if quantum 
correlations are present in the initial state 
\cite{Partovi,Micadei}.
\subsubsection{Paradigm 1}
Consistently with the classical case, we define the temperature of a 
two-level system as the derivative of the von Neumann entropy with 
respect to energy in a zero work process, a condition which is satisfied 
in the standard framework if the Hamiltonian is time-independent. 
Since the Hamiltonian is determined by the effective magnetic field 
$\vec{v}=\varepsilon\hat{v}$, fixing the direction $\hat{v}$ we define   
\begin{equation}
\dfrac{1}{\mathcal{T}}=\dfrac{\partial S_{vN}}{\partial E} 
\biggr{\vert}_{\varepsilon}.
\end{equation}
We observe that $S_{vN}$ depends only on $B$, which, in turn, depends 
on three arbitrary orthogonal components of the Bloch vector. 
Due to Eq. (\ref{E1}), the energy depends only on the component 
parallel to $\hat{v}$,
\begin{equation}\label{E2}
E=-\varepsilon\vec{B}.\hat{v},
\end{equation} 
\noindent so we have:
\begin{equation}
\dfrac{1}{\mathcal{T}}=\dfrac{dS_{vN}}{dB}\dfrac{\partial B}
{\partial {(\vec{B}.\hat{v})}}\dfrac{\partial (\vec{B}.\hat{v})}
{\partial E}\biggr{\vert}_{\varepsilon}.
\end{equation}
\noindent From Eq. (\ref{entropy1}),
\begin{equation}\label{dSvN}
\dfrac{dS_{vN}}{dB}=-k_{B}\tanh^{-1}(B),
\end{equation}
and the other factors are
\begin{equation}
\dfrac{\partial B}{\partial {(\vec{B}.\hat{v})}}=\hat{B}.\hat{v};\hspace{0.2cm}
\dfrac{\partial (\vec{B}.\hat{v})}{\partial E}\biggr{\vert}_{\varepsilon}=-\dfrac{1}{\varepsilon}.
\end{equation}
\noindent Thus,
\begin{equation}\label{TP1}
\mathcal{T}=\dfrac{\varepsilon}{k_{B}(\hat{B}.\hat{v})\tanh^{-1}(B)}.
\end{equation} 
We notice that pure states have zero temperature, except for those such 
that the Bloch vector is orthogonal to the effective magnetic field, 
for which the temperature is not defined. 
In the case of mixed states, the temperature diverges as the 
magnetization in the direction of $\hat{v}$ goes to zero. 
This behavior is similar to that corresponding to classical spin 
systems \cite{Salinas}.
 
\subsubsection{Paradigm 2}
From this new perspective, a zero-work process is implemented 
keeping constant the product $\hat{B}.\vec{v}$, so we define
\begin{equation}
\dfrac{1}{T}=\dfrac{\partial S_{vN}}{\partial E} 
\biggr{\vert}_{\hat{B}.\vec{v}},
\end{equation}
\noindent or, equivalently,
\begin{equation}\label{TP22}
\dfrac{1}{T}=\dfrac{dS_{vN}}{dB}\dfrac{\partial B}
{\partial E}\biggr{\vert}_{\hat{B}.\vec{v}}
\end{equation}
\noindent In this case it is convenient to write Eq. (\ref{E1}) 
in the form
\begin{equation}\label{B(E)}
B=-\dfrac{E}{\hat{B}.\vec{v}},
\end{equation}
\noindent 
\noindent so that the second factor in Eq. (\ref{TP22}) is written
\begin{equation}
\dfrac{\partial B}{\partial E}\biggr{\vert}_{\hat{B}.\vec{v}}=-\dfrac{1}{\hat{B}.\vec{v}},
\end{equation}
\noindent and, therefore,
\begin{equation}\label{TP2}
T=\dfrac{\varepsilon\hat{B}.\hat{v}}{k_{B}\tanh^{-1}(B)}.
\end{equation}
Two families of zero-temperature states appear in paradigm 2: 
those with $B=1$ (pure states), and $\hat{B}.\vec{v}=0$ (SMCS states). 
On the other hand, the only infinite-temperature state is the 
maximally mixed state.  

\subsubsection{Discussion}

First, we note the different position of the factor $\hat{B}.\hat{v}$ 
in Eqs. (\ref{TP1}) (first paradigm) and in Eq. (\ref{TP2}) (second paradigm). 
This implies that the relation between both temperatures is
\begin{equation}\label{TTrelation}
T=\mathcal{T}(\hat{B}.\hat{v})^{2}=\mathcal{T}\cos^{2}\theta ,
\end{equation}
\noindent from which we deduce that they have always the same 
sign and that, for all possible states, $T\leq\mathcal{T}$. 
In particular, for incoherent states in the energy eigenbasis 
$(\theta =0)$ both temperatures coincide:
\begin{equation}\label{Tincoh}
T=\mathcal{T}=\dfrac{\varepsilon}{k_{B}\tanh^{-1}(B)}
\end{equation} 
If additionally the system reaches thermal equilibrium with an 
environment at temperature $T_{E}$, the reduced state is described 
by the Gibbs state in which the populations of the ground and the 
excited levels $P_{g}^{eq}$ and $P_{e}^{eq}$ are fixed by the 
environment temperature \cite{Pathria},
\begin{equation}\label{Teq}
T_{E}=\dfrac{2\varepsilon}{k_{B}\ln{\left(\dfrac{P_{g}^{eq}}
		{P_{e}^{eq}}\right)}}
\end{equation}
\noindent Since in this case the Hamiltonian and the density matrix 
commute, the populations and the eigenvalues of the density matrix 
coincide.
Therefore, from Eq. (\ref{eigenvalues}), we obtain,
\begin{equation}\label{pgpe}
\ln{\left(\dfrac{P_{g}^{eq}}{P_{e}^{eq}}\right)}=\ln{\left(\dfrac{1+B^{eq}}{1-B^{eq}}\right)}=2\tanh^{-1}(B^{eq}).
\end{equation}
\noindent Finally, from Eqs. (\ref{Tincoh}), (\ref{Teq}) and (\ref{pgpe}), 
we conclude that, in thermal equilibrium,
\begin{equation}
T=\mathcal{T}=T_{E}.
\end{equation}
\noindent Therefore, at least in principle, both expressions 
(\ref{TP1}) and (\ref{TP2}) extend naturally the concept of 
temperature to the non-equilibrium situation.

We have already noted that in the context of paradigm 2, 
the zero-energy plane divides the Bloch sphere in two hemispheres 
with opposite values of temperature. This is also true in paradigm 1, 
since both temperatures have the same sign. 
This can be seen explicitly finding the energy-temperature relation 
from Eqs. (\ref{E2}) and (\ref{TP2}),
\begin{equation}
T=-\dfrac{E}{k_{B}B\tanh^{-1}(B)}
\end{equation}
Finally, we note that in paradigm 1, zero temperature implies 
zero entropy. This is not true in paradigm 2, since SMCS states 
have zero temperature but its entropy may take any value.

\begin{figure}
\centering
\begin{minipage}{0.5\columnwidth}
  \centering
  \includegraphics[trim= 300 50 0 0,scale=0.3,clip]{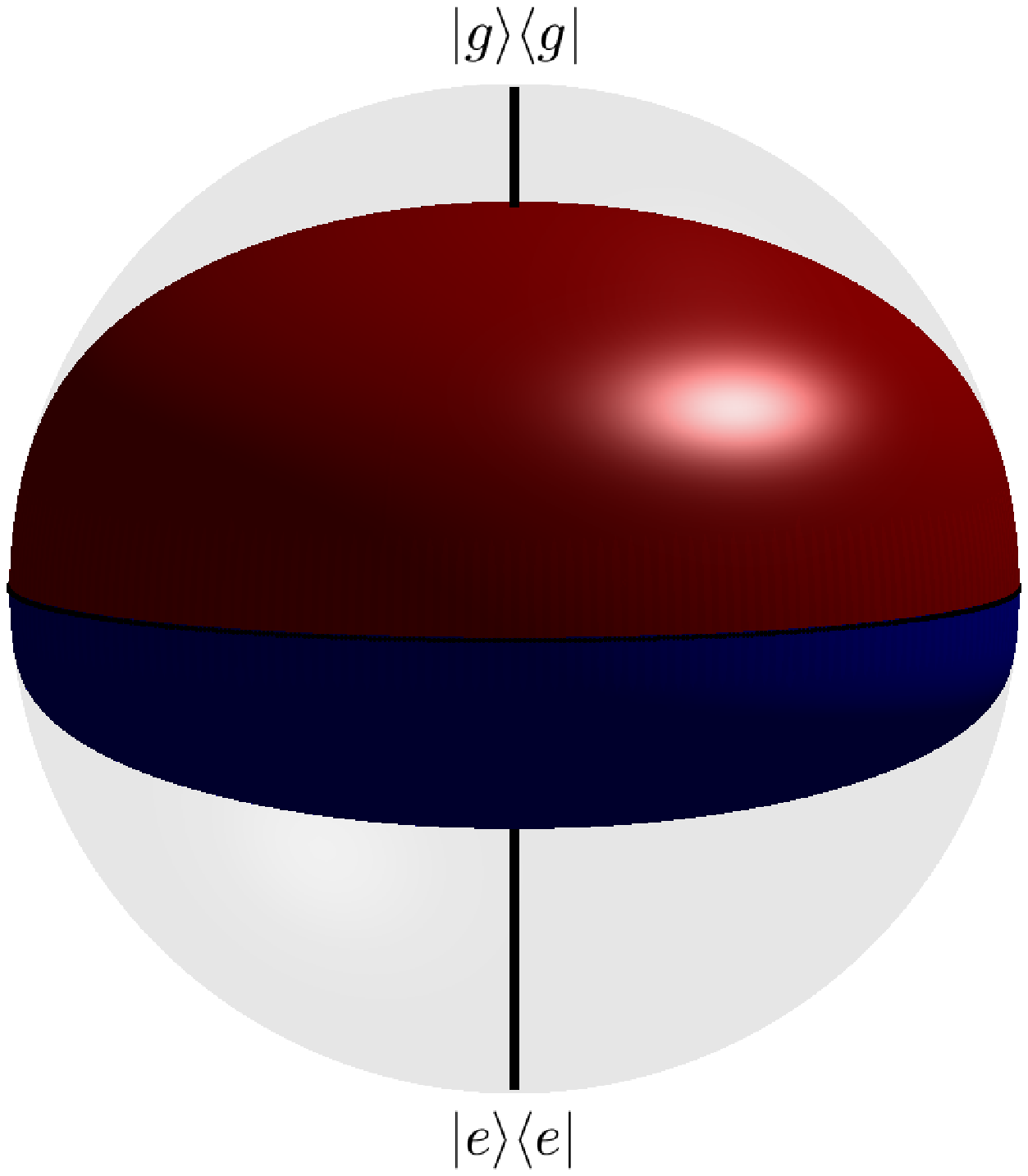}
\end{minipage}%
\begin{minipage}{0.5\columnwidth}
  \centering
  \includegraphics[trim= 340 50 -50 0,scale=0.3,clip]{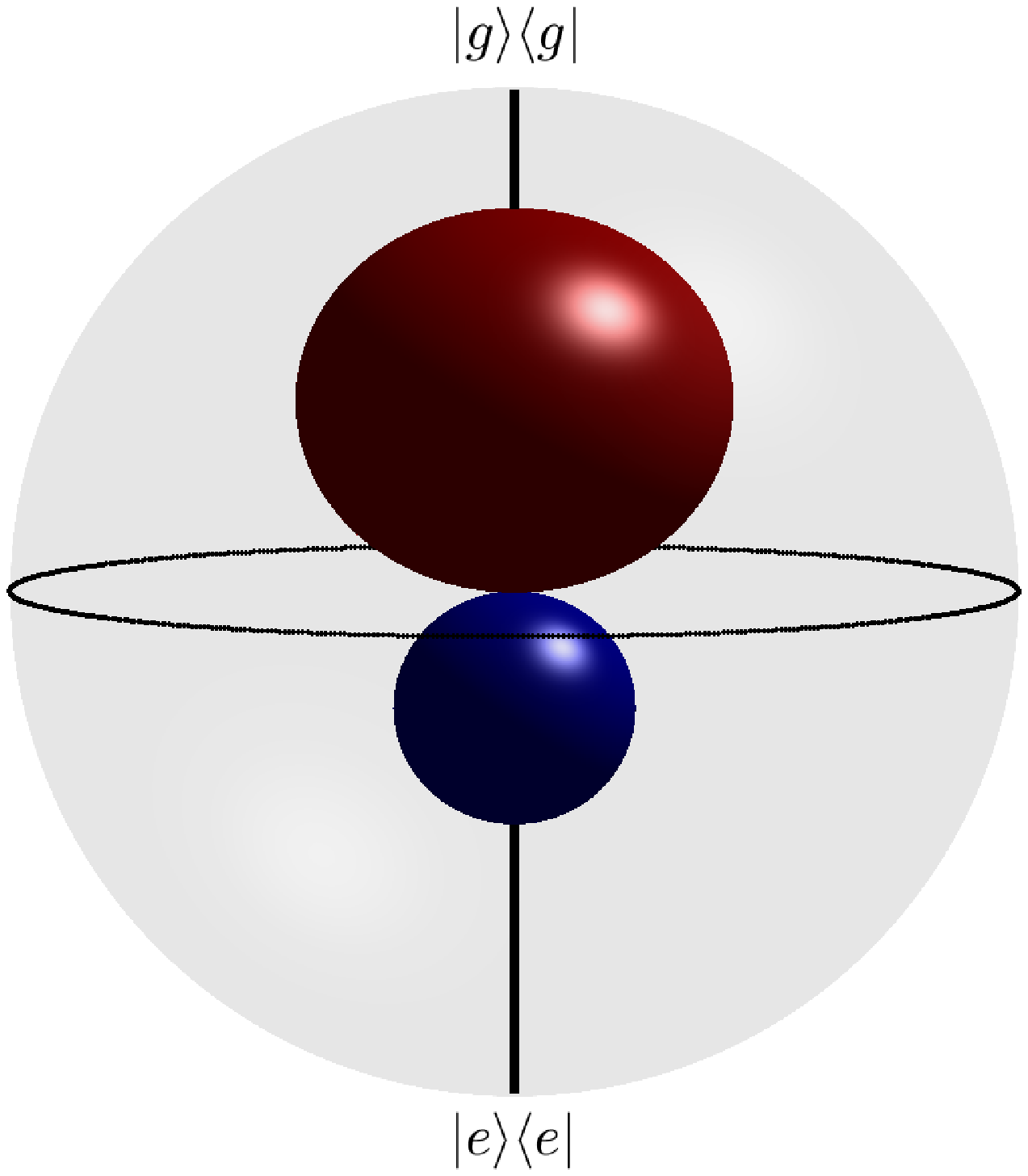}
\end{minipage}
\caption{Isothermal surfaces in the Bloch sphere, corresponding to the
temperature values $k_{B}T_{1}=\varepsilon$ (red, upper region) 
and $k_{B}T_{2}=-2\varepsilon$ (blue, lower region) in paradigm 1 
(left panel) and paradigm 2 (right panel).}
\label{f:f1}
\end{figure}

Some constant temperature surfaces in both paradigms are shown 
in Fig. (\ref{f:f1}).

\subsection{Heat capacity} 
As usual, we define the heat capacity as the partial derivative of the 
energy with respect to temperature, in a zero-work process. 
\subsubsection{Paradigm 1}
The derivation in Alicki's theoretical framework was performed in 
Ref. \cite{Vallejo1},
\begin{small}
\begin{equation}\label{CvP1}
\mathcal{C}_{\varepsilon}=\frac{k_{B}B(1-B^{2})
	[\tanh^{-1}(B)]^{2}(\vec{B}.\hat{v})^{2}}
{\tanh^{-1}(B)(B^2-(\vec{B}.\hat{v})^2)(1-B^2)+B(\vec{B}.\hat{v})}.
\end{equation}
\end{small}
\noindent where the consequences of this result are discussed in detail. 
\subsubsection{Paradigm 2}
In the new approach, the heat capacity is
\begin{equation}
C_{\hat{B}.\vec{v}}=\dfrac{\partial E}{\partial T}
\biggr{\vert}_{\hat{B}.\vec{v}}.
\end{equation}
The evaluation of the above equation requires to express the energy in 
terms of $T$ and $\hat{B}.\vec{v}$. 
From Eq. (\ref{TP2}):
\begin{equation}
B=\tanh\left(\dfrac{\hat{B}.\vec{v}}{k_{B}T}\right),
\end{equation}
\noindent which, combined with Eq. (\ref{EP2}), leads to
\begin{equation}
E=-(\hat{B}.\vec{v})\tanh\left(\dfrac{\hat{B}.\vec{v}}{k_{B}T}\right).
\end{equation}
\noindent Therefore,
\begin{equation}\label{CvP2}
C_{\hat{B}.\vec{v}}=k_{B}\left[\dfrac{x}{\cosh(x)}\right]^{2},
\end{equation}
\noindent where 
\begin{equation}
x=\dfrac{\hat{B}.\vec{v}}{k_{B}T},
\end{equation}
\noindent and $T$ is given by Eq. (\ref{TP2}).
\subsubsection{Discussion}
A quick inspection of Eqs. (\ref{CvP1}) and (\ref{CvP2}) shows that in both cases the heat capacity is non-negative for all possible states. 
The equivalence between Eq. (\ref{CvP1}) and the equilibirium heat capacity is shown in Ref. \cite{Vallejo1}. 
On the other hand, the expression (\ref{CvP2}) for the heat capacity 
in the new paradigm is clearly a more natural extension of the 
classical result. 
Since in thermal equilibrium the Bloch vector is parallel 
to the effective magnetic field,
\begin{equation}
\hat{B}.\vec{v}=\vert\vec{v}\vert=\varepsilon,
\end{equation}
\noindent and since in that case the temperature of the system  
equals the environment temperature, Eq. (\ref{CvP2}) reduces to 
the well-known expression \cite{Mahdavifar}:
\begin{equation}
C_{\hat{B}.\vec{v}}=k_{B}\left[\dfrac{\varepsilon/k_{B}T}
{\cosh\left(\varepsilon/k_{B}T\right)}\right]^{2}.
\end{equation}
\subsection{Entropy production}
Classically, the entropy change of a closed system is given by
\begin{equation}\label{2ndlaw}
dS=\dfrac{\delta Q}{T}+\delta S_{gen}^{int},
\end{equation}
\noindent where the first term corresponds to the entropy flux 
through the system's boundary at temperature $T$ due to 
heat exchange, and the second term is the non-negative entropy 
production associated to the irreversibilities inside the 
system \cite{Cengel}. 
A typical situation is that in which the system is in contact 
with a heat bath at temperature $T_{E}$. 
In this case, if the system's temperature and the environment temperature are different, an additional entropy production appears 
due to the irreversible character of the heat transfer. 
In this case, the total entropy production can be evaluated by 
applying Eq. (\ref{2ndlaw}) to the system plus its border, so that the 
irreversible heat transfer occurs in its interior, 
i.e., considering the environment temperature instead of the 
system's temperature in Eq. (\ref{2ndlaw}).
In this case, 
\begin{equation}\label{dSgentot}
dS=\dfrac{\delta Q}{T_{E}}+\delta{S_{gen}^{tot}}.
\end{equation} 

From the above equations we can make two important observations. 
One is that the total entropy production can be separated in 
the internal and the heat transfer contributions. 
The latter corresponds to the second term in the right-hand 
side of
\begin{equation}\label{dsgeninttot}
\delta{S_{gen}^{tot}}=\delta S_{gen}^{int}+
\delta Q\left(\dfrac{1}{T}-\dfrac{1}{T_{E}}\right).
\end{equation}
We also note that Eq. (\ref{dSgentot}) can be written as
\begin{equation}
\delta{S_{gen}^{tot}}=dS-\dfrac{\delta Q}{T_{E}}.
\end{equation}
\noindent Note that the two terms on the right-hand side of 
this equation correspond to the entropy variations of the system 
and the environment respectively, so the total entropy production 
and the total entropy variation, in the case of classical 
systems, coincide: 
\begin{equation}
\delta{S_{gen}^{tot}}=dS^{tot}.
\end{equation}

This last equation has represented a big challenge to the 
possibility of extending Thermodynamics to the quantum regime. 
Since the evolution of an open system is, in the general case, 
irreversible, one expects a positive entropy production, 
($\delta S_{gen}^{tot}>0$). 
However, the unitary evolution of the whole system preserves 
the density matrix eigenvalues, and, as a consequence the 
total entropy does not change ($dS^{tot}=0$). 
This fundamental problem has been addressed in several works 
\cite{Deffner,Breuer, Esposito,Brunelli}, and it has been suggested 
that entropy production instead of entropy change is the relevant 
quantity in order to explain irreversible behavior. 
In this work, we will only focus in the analysis of Eq. (\ref{2ndlaw}) 
within each paradigm, in order to investigate if an intrinsic 
entropy production is expected in each case.
\subsubsection{Paradigm 1}  
This problem has been analyzed in Ref. \cite{Vallejo1}. 
From Eqs. (\ref{heatP1}), (\ref{dSvN}) and  (\ref{TP1}), it is straightforward 
to obtain an equation linking the von Neumann entropy, 
the heat transferred, and the temperature defined in Alicki's 
theoretical framework:
\begin{equation}
dS_{vN}=\dfrac{\delta \mathcal{Q}}{\mathcal{T}}+
\delta \mathcal{S}_{gen}^{int},
\end{equation}
\noindent where the internal entropy production is given by 
\begin{equation}\label{sgenP1}
\delta\mathcal{S}_{gen}^{int}=
-k_{B}\tanh^{-1}(B)[\hat{B}-(\hat{v}.\hat{B})\hat{v}].d\vec{B}.
\end{equation}
\noindent Since $\hat{B}-(\hat{v}.\hat{B})\hat{v}$ is orthogonal 
to $\hat{v}$, for an unitary evolution, or if the system evolves 
along equilibrium states, Eq. (\ref{sgenP1}) predicts zero internal 
entropy production, as expected.
\subsubsection{Paradigm 2}
From Eqs. (\ref{heatP2}), (\ref{dSvN}) and  (\ref{TP2}), 
\begin{equation}
dS_{vN}=\dfrac{\delta Q}{T},
\end{equation}
\noindent and, as a consequence,
\begin{equation}
\delta S_{gen}^{int}=0.
\end{equation}
\noindent Thus, in the case of paradigm 2, no intrinsic entropy 
production is expected in any process. 
It must be pointed out that this result is valid only in 
the case of two-level systems. 
For higher dimensional systems we have been able to find a very 
reduced set of quantum states for which the concept of local 
temperature can be consistently defined, but
not a generally valid treatment of this quantity \cite{Vallejo2}.

\subsubsection{Discussion}
To give a physical interpretation to Eq. (\ref{sgenP1}), 
we first recall the definition of heat in paradigm 1, Eq. (\ref{heat1}),
$$\delta \mathcal{Q}=-d\vec{B}.\vec{v},$$
\noindent and note that only the part of $d\vec{B}$ which is 
parallel to $\hat{v}$ is responsible for heat exchange. 
If we restrict ourselves to the case in which the Hamiltonian is fixed, 
expressing Eq. (\ref{sgenP1}) in spherical coordinates (with the 
$z$ axis in the direction of $\hat{v}$), we obtain \cite{Vallejo1}
\begin{equation}\label{sgenP12}
\delta S_{gen}^{int}=-k_{B}\tanh^{-1}(B)\sin{\theta}d(B\sin\theta).
\end{equation} 
\noindent Therefore, the component of $d\vec{B}$ which is orthogonal 
to $\hat{v}$ and produces no heat, is the one responsible for  
entropy production. 
Since $B\sin\theta=C_{l_{1}}$, the entropy produced in paradigm 1 
is proportional to the change in the coherence of the qubit in the 
energy eigenbasis. 
If coherence is lost, entropy is produced, and destruction of 
entropy can occur in processes in which the coherence of the 
qubit increases.
On the other hand, the non-existence of internally generated entropy 
in paradigm 2 was expected, since in it heat is defined as the part 
of the energy change which produces an entropy change. 
Therefore, in this approach, there are no causes of entropy 
variation other than the heat flow.

The non-existence of internal entropy production is consistent with 
the possibility of obtaining an extra amount of work in comparison 
with the previous approach. 
Nevertheless, if the system and its environment are at different 
temperatures, entropy production at the boundary should be 
expected due to irreversible heat transfer
according to Eq. (\ref{dsgeninttot}),
\begin{equation}
\delta S_{gen}^{ht}=\delta Q\left(\dfrac{1}{T}-\dfrac{1}{T_{E}}\right).
\end{equation}
\noindent However, this result was obtained by subtracting 
Eqs. (\ref{2ndlaw}) and ({\ref{dSgentot}}), and it was assumed 
that the heat released by one system equals the one absorbed 
by the other, an aspect that in the quantum case is not guaranteed 
in either of the two paradigms. 
In fact, for a system of two qubits in positive-temperature states, 
and under a global unitary evolution, the Schmidt decomposition 
forces both entropy changes to be equal. 
As a consequence, in paradigm 2 the heat exchanged has the same 
sign for both systems, so they are releasing or absorbing heat simultaneously.

\section{Examples}
\subsection{Two-level atom in a heat bath}

In the Markovian approximation (valid in the high temperature 
limit), the evolution of a two-level atom interacting with a 
thermal state of the electromagnetic field at temperature $T_{E}$ 
is given, in the interaction picture, by the master equation
\cite{Breuer}

\begin{align}
{\frac{\partial\rho_{_{S}}}{\partial t}}&=
\gamma_0(\mathcal{N}+1)\left(\sigma_{_-}\rho\sigma_{_+}-
\frac{1}{2}\sigma_{_+}\sigma_{_-}\rho-
\frac{1}{2}\rho\sigma_{_+}\sigma_{_-}\right)\,\notag \\
+&\gamma_0\mathcal{N}\left(\sigma_{_+}\rho\sigma_{_-}-
\frac{1}{2}\sigma_{_-}\sigma_{_+}\rho-
\frac{1}{2}\rho\sigma_{_-}\sigma_{_+}\right) ,\, 
\label{MasterEq1}
\end{align}
\noindent where $\gamma_{0}$ is the spontaneous emission rate, 
$\omega_{0}$ is the transition frequency, 
$\mathcal{N}$ is the Planck distribution at that frequency,
\begin{equation}
\mathcal{N}=\frac{1}{e^{\beta_{E}\hbar\omega_{0}}-1} ,
\end{equation}
\noindent $ \sigma_{\pm}=\frac{1}{2} (\sigma_{x}\pm i\sigma_{y})$, and $\beta_{E}=(k_{B}T_{E})^{-1}$.

\begin{figure}[!h]
 \centering
    \includegraphics[width=0.75\columnwidth]{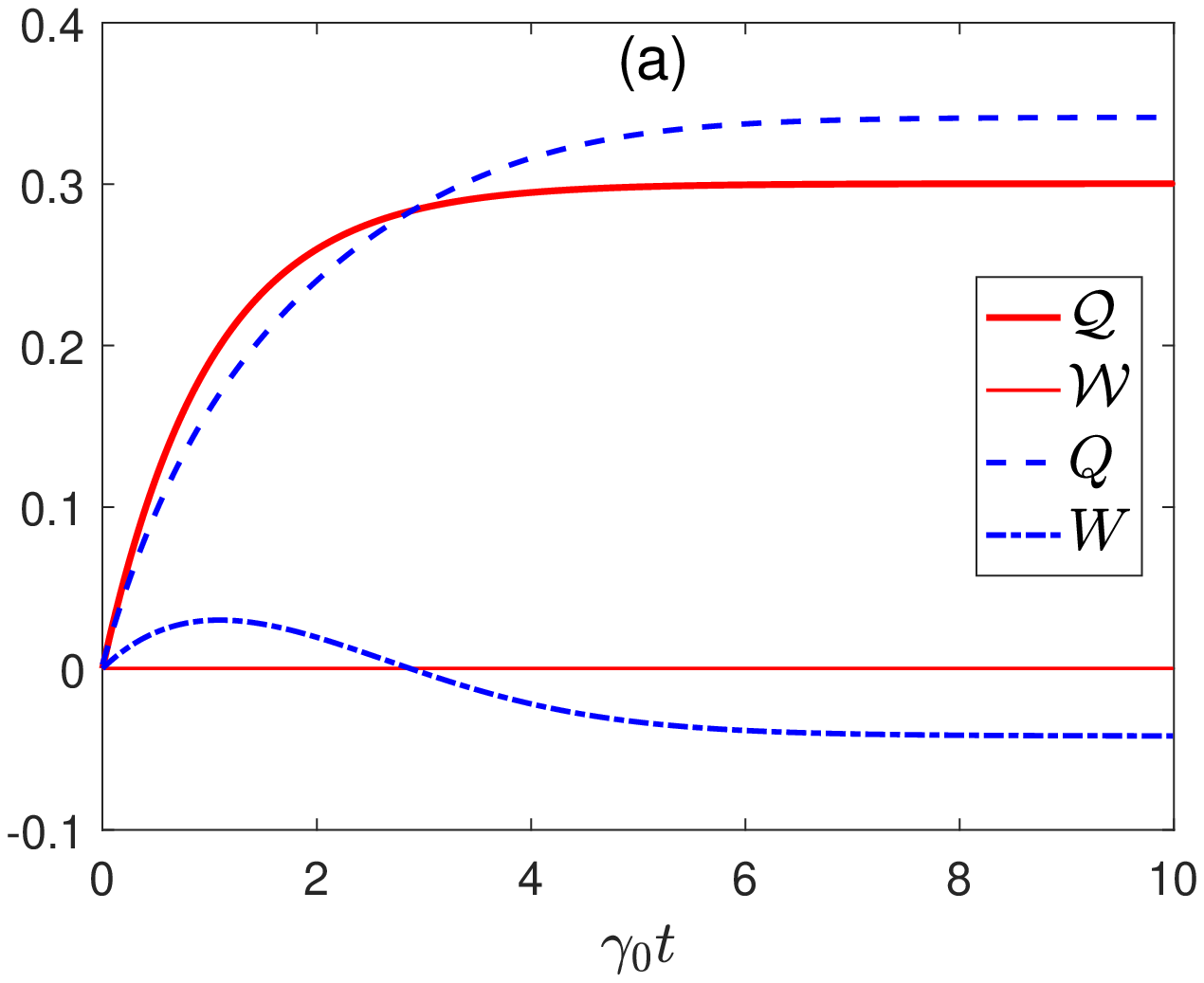}
    \includegraphics[width=0.75\columnwidth]{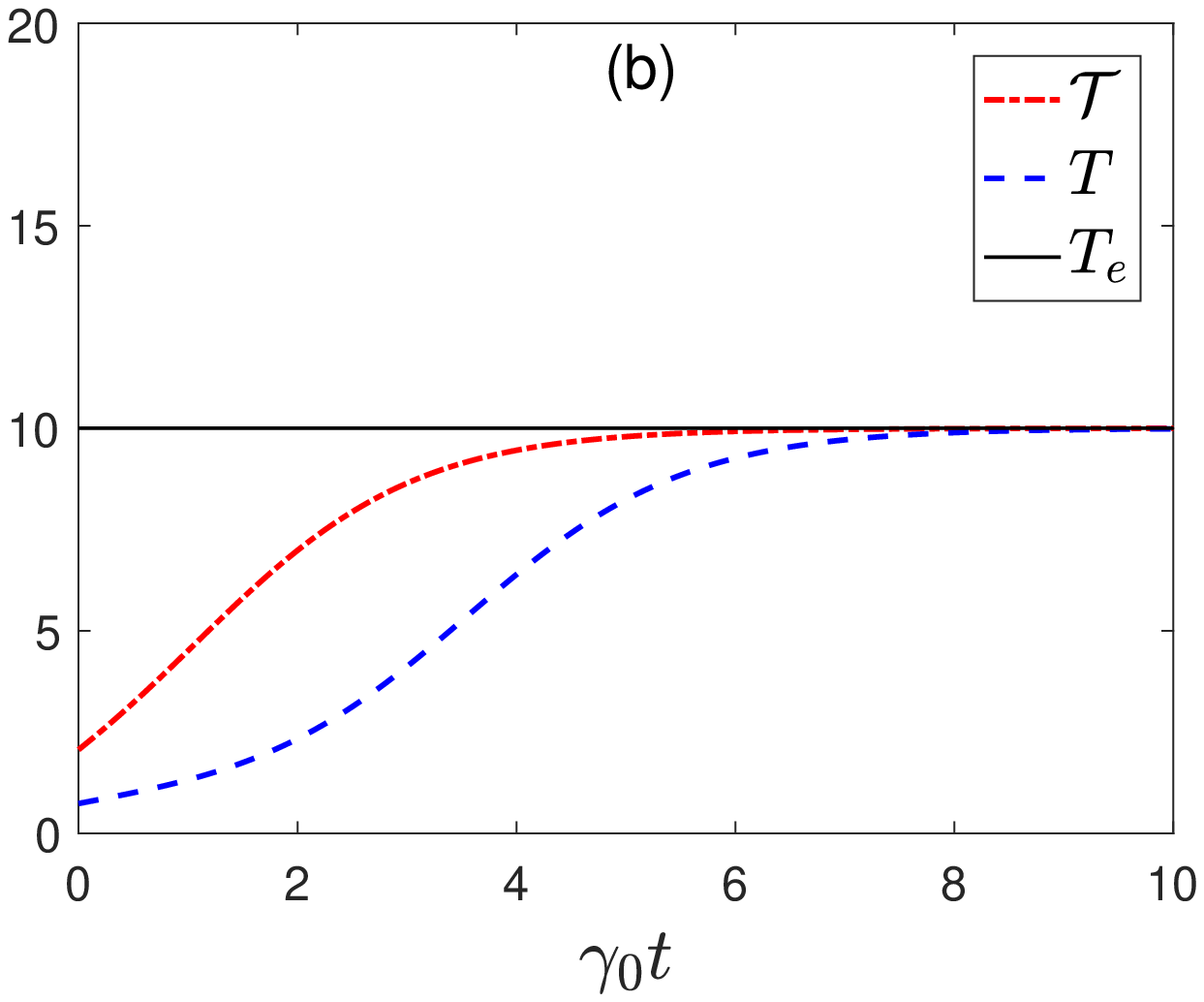}
 \caption{Comparative evolution of the thermodynamic quantities 
 in both paradigms, for  two-level atom interacting with a thermal 
 electromagnetic field at temperature $k_{B}T_{e}/\varepsilon=10$.
 In panel (a) we show the heat and work exchanged, 
 in panel (b) the temperature.
 In both cases the initial state is a product state, where the 
 atom's initial density matrix is defined by the Bloch vector 
 $\vec{B}=(0.2,0.5,0.4)$.}
\label{f:f2}
\end{figure}

It is known that in the asymptotic regime, the equilibrium state 
of the atom is described by the thermal reduced density matrix,
\begin{equation}\label{thermal state}
\rho^{eq}=\dfrac{e^{-\beta_{E} H}}
{\text{tr}\left( e^{-\beta_{E} H}\right)},
\end{equation}
\noindent which implies that the Bloch vector points in the 
direction of the effective magnetic field, with modulus
\begin{equation}
B^{eq}=\tanh(\beta_{E}\varepsilon),
\end{equation}
\noindent where $\varepsilon=\vert\vec{v}\vert$ is the eigenenergy 
of the system. 
As a consequence, the environment temperature determines the equilibrium 
values of all the thermodynamic quantities. 
In particular, the equilibrium temperature coincides with the 
environment temperature. 

In the case of paradigm 1, the total energy variation of the atom 
corresponds to the heat exchanged with the environment, 
$\Delta E=\mathcal{Q}$, represented by the thick red continuous 
line in Fig. \ref{f:f2}(a). 

From the new perspective of paradigm 2, the thermalization process 
is related to two different phenomena.  
On the one hand, since the initial entropy of the atom is arbitrary 
and its final entropy is defined by the environment temperature, 
a heat exchange is needed so that the final entropy is the one that 
ensures thermal equilibrium with the environment.
On the other hand, work is also required to rotate the Bloch 
vector towards the equilibrium direction. 
Both quantities are also represented in Fig. \ref{f:f2}(a).

\begin{figure}[!h]
\centering
  \includegraphics[trim= 340 50 -50 0, scale=0.6,clip] {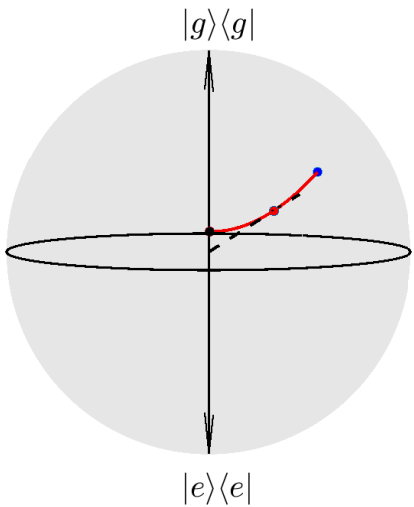}
\caption{Evolution of the state of the atom in the Bloch sphere. 
The system evolves from the initial state (0.2,0.5,0.4) (blue point) towards the thermal state (red point) located at the vertical diameter.}
 \label{f:f3}
\end{figure}

The transient positive character of the work observed in the case of
paradigm 2 can be understood by analyzing the path towards the equilibrium state in the Bloch sphere, see Fig. \ref{f:f3}.
Note that when evolution begins, even though both the distance 
to the $z$ axis, $B\sin\theta$ and $B$ decrease, 
$\theta$ increases, so work is done on the system. 
This occurs until the point representing the reduced state reaches 
the intersection of the trajectory with the tangent  line from the center of the sphere (the dashed line in Fig. \ref{f:f3}). 
From that point onwards, both $B$ and $\theta$, and consequently 
the net work, begin to decrease, resulting, at the end of the 
thermalization process, in a total negative work done on the 
system . 
For this reason, thermalization in the case of paradigm 2 requires 
greater heat absorption from the environment, part of which is 
converted into work. 

The first part of the process described above shows that 
although coherence is a useful resource, its consumption 
does not necessarily imply an extraction of work. 
In fact, if the temperature of the bath is infinite, 
there are trajectories that converge to the maximally mixed 
state tangentially to the plane $z=0$, so a positive total 
work is performed on the system during the process.

Regarding the behavior of the temperature, in Fig. \ref{f:f2}(b) 
it is shown that in both theoretical frameworks the temperature 
increases as the atom absorbs heat. 
As expected, the respective temperatures tend to the equilibrium temperature, with a faster convergence in Alicki's formulation.

\subsection{Photon exchange between two two-level atoms}

As a second example, let us consider a system composed by two 
two-level atoms embedded in a common environment at zero temperature. 
If the atoms are separated a distance $R$ and only the spontaneous 
emission is taken into account, the system undergoes a dissipative 
process described by the master equation \cite{Jakobczyk}
\begin{equation}\label{MastEQ1}
\frac{\partial \rho}{\partial t}
=\frac{1}{2}\sum_{k,l=A,B}\gamma_{kl}\left(2\sigma_{-}^{k}
\rho\sigma_{+}^{l}
-\sigma_{+}^{k}\sigma_{-}^{l}\rho-\rho\sigma_{+}^{k}\sigma_{-}^{l}\right),
\end{equation}
\noindent where
\begin{equation}
\sigma_{\pm}^{A}=\sigma_{\pm}\otimes\mathbb{I}_{2},
\hspace{0.2cm}\sigma_{\pm}^{B}=\mathbb{I}_{2}\otimes\sigma_{\pm} ,
\end{equation}
\noindent $\gamma_{AA}=\gamma_{BB}=\gamma_{0}$ is the
spontaneous emission rate of each atom, 
$\gamma_{AB}=\gamma_{BA}=\gamma=g(R)\gamma_{0}\leq\gamma_{0}$ 
is the photon-exchange relaxation constant, and  
$g(R)$ is a function which approaches the value 1 as $R\rightarrow 0$. 

In the case $\gamma <\gamma_{0}$, the atoms are not capable of 
absorbing all the energy emitted by the other, so independently 
of the initial state, the composed system asymptotically relaxes 
towards the ground state $\ket{0}\otimes\ket{0}$.

Let us analyze the case in which the system starts from an 
initially uncorrelated state, with the atom (A) in the partially 
excited state defined by the Bloch vector 
$\vec{B}_{A}=(0,0.5,0.8)$, 
while atom B is in the ground state, $\vec{B}_{B}=(0,0,1)$. 

\begin{figure}[!h]
 \centering
 \includegraphics[width=0.7\columnwidth]{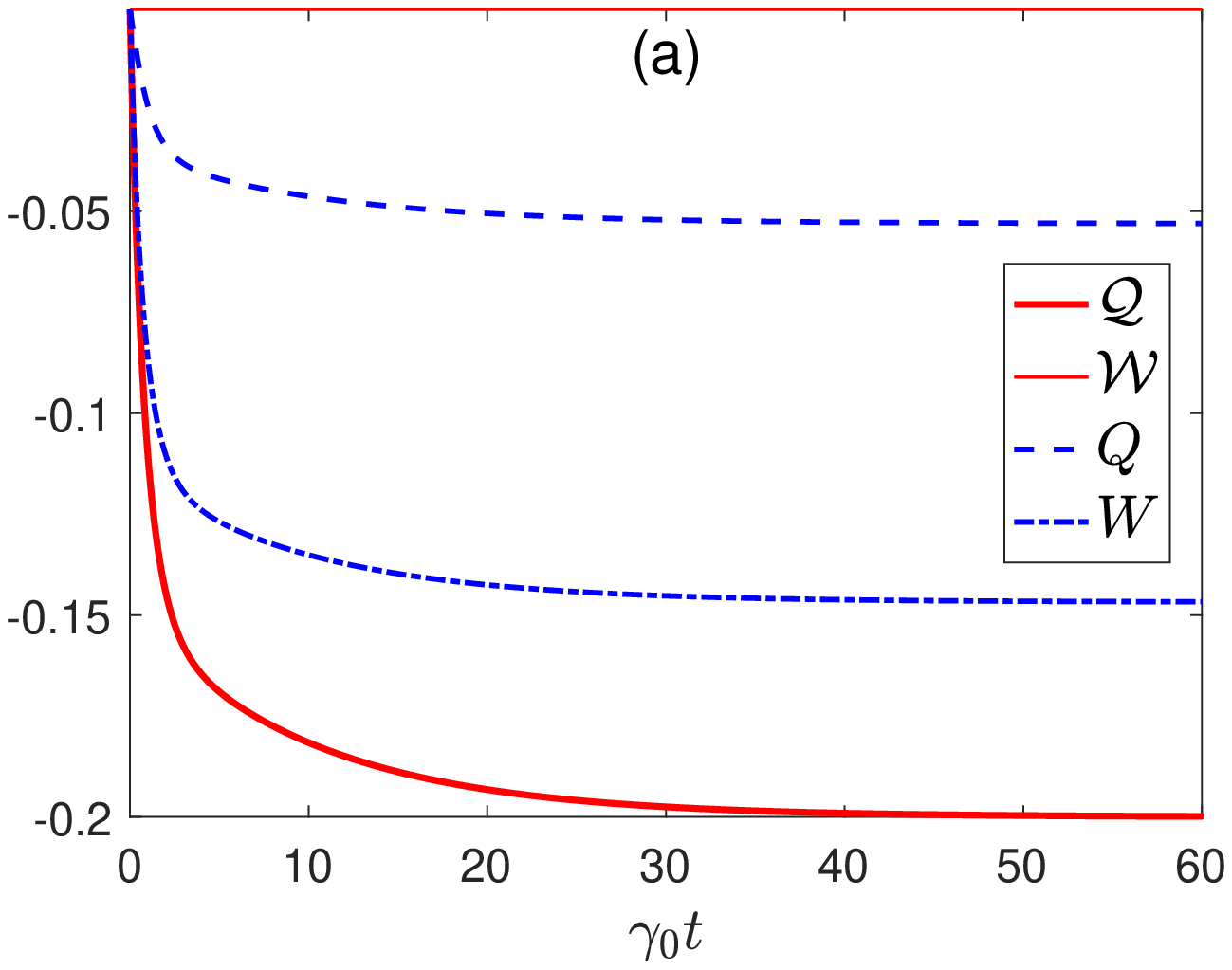}
 \includegraphics[width=0.7\columnwidth]{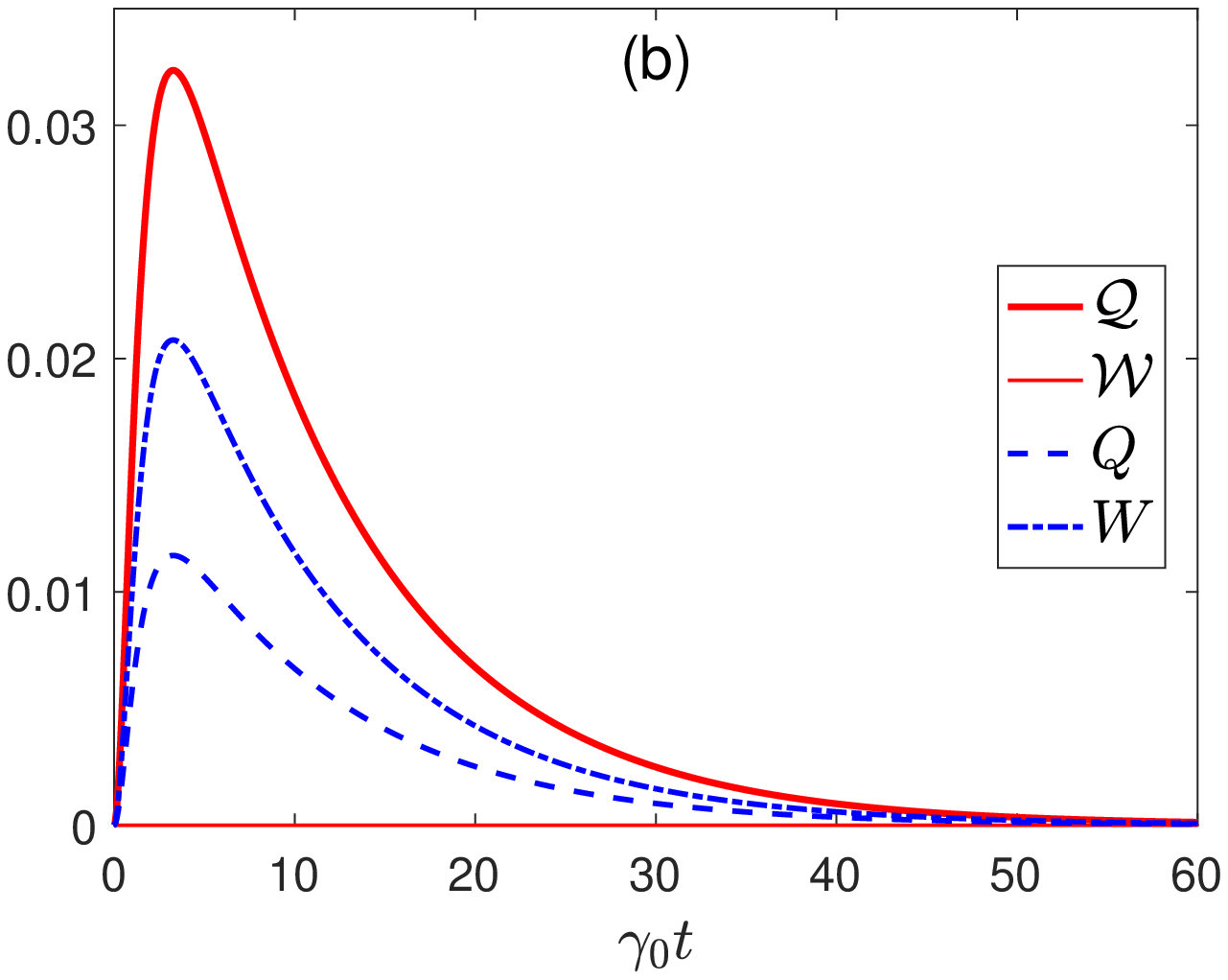}
 \includegraphics[width=0.7\columnwidth]{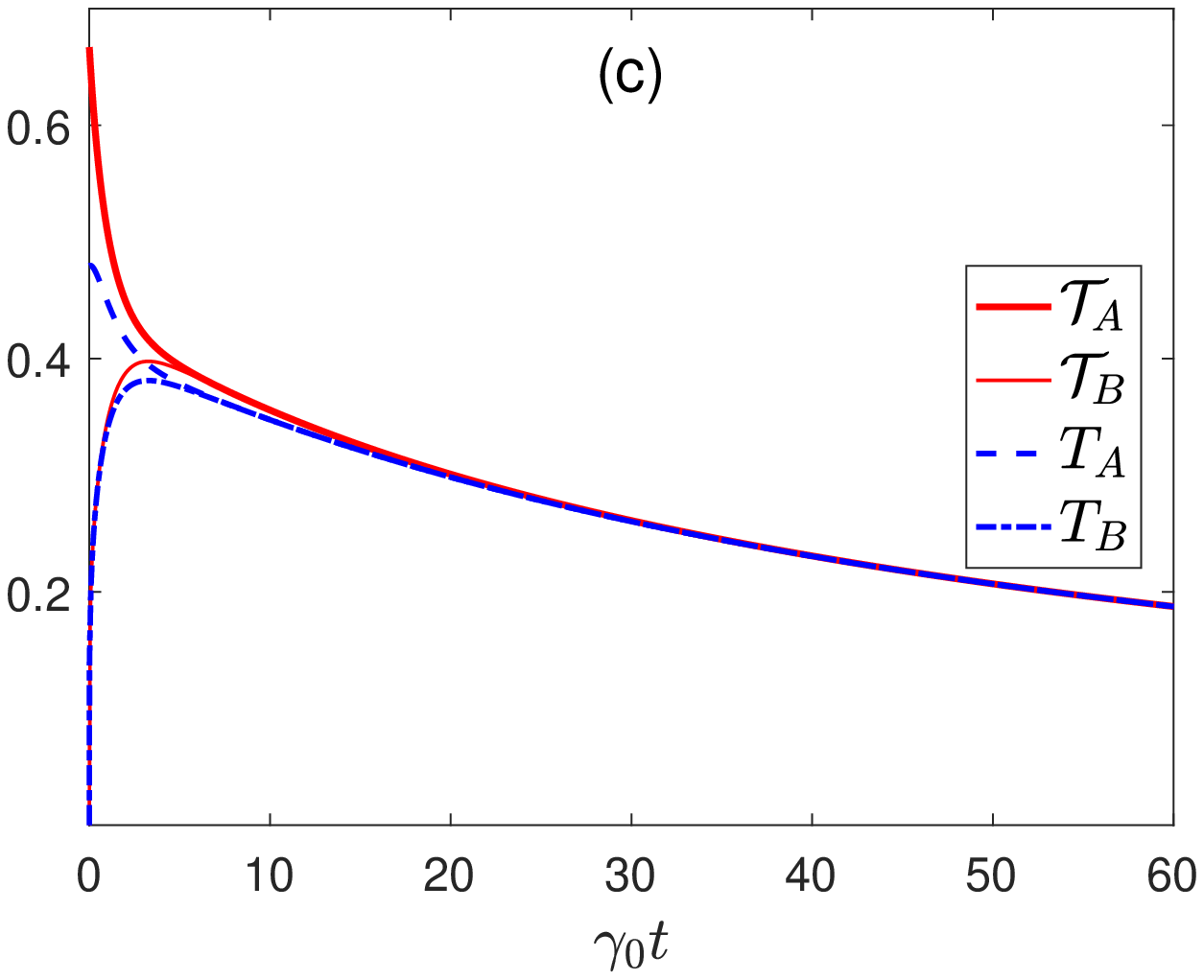}   \caption{Thermodynamic quantities for the evolutions of the two atoms considered in Fig. \ref{f:f4} using the two paradigms considered in 
 this work. 
 In panel (a) are shown the heat and work exchanged by atom A.
 Panel (b) shows the same quantities for atom B. 
 The respective temperatures, also in both approaches, are shown 
 inpanel (c) .} 
 \label{f:f5}
\end{figure}

Since the local Hamiltonians are constant in time, from the point 
of view of paradigm 1 the emission and absorption of photons is 
modeled as a heat transfer process between the atoms, with some 
heat released to the environment. 
The heat exchanged by each atom is represented by the thick continuous 
red lines in Figs. \ref{f:f5}(a) and \ref{f:f5}(b).
We note that in the net balance, the atom A is always releasing heat 
but at a decreasing rate as it approaches the ground state, while 
atom B undergoes the process described in the previous paragraph, 
interpreting its energy change exclusively as heat absorbed and 
released.

From the perspective of paradigm 2, the energy variation of atom A 
includes a negative work component, i.e. work performed by the system, 
represented by the dash-dot blue line in Fig. \ref{f:f5}(a).
This is due to the fact that the angle formed by the Bloch vector 
and the vertical direction decreases monotonically in time as  
the atom approaches the ground state. 
As a consequence, the amount of heat emitted by the atom is less 
than in paradigm 1. 

Regarding atom 2, we note that in the first part of the evolution 
the photon absorption has two effects: it leads to an increase in 
entropy, interpreted as heat entering the system.
In fact, we note in Fig. \ref{f:f5}(c) that the temperature also increases. 
Besides, it also leads to a change in the direction of the 
Bloch vector, which moves away from the vertical direction.
This motion requires external rotational work to overcome its 
tendency to stay in the vertical direction due 
to the presence of the magnetic field. 
We also note that the signs of heat and work are in phase.

As the system subsequently evolves from state $2B$ to $3B$, the 
emission of photons governing the process comes from the hand of 
heat released and work done, in amounts opposite to those of the 
process $1B \rightarrow 2B$, with a decrease in temperature.
It is also interesting to note that the state $2B$ occurs 
approximately when the temperatures of the atoms are equal, 
so it is reasonable to think that, from that moment on, 
the energy flows between the atoms balance, and both atoms cool releasing energy to the environment, as it can be seen in 
Fig. \ref{f:f5}.
 
\begin{figure}[!h] 
	\centering
	\includegraphics[trim= 340 50 -50 0, scale=0.6,clip] {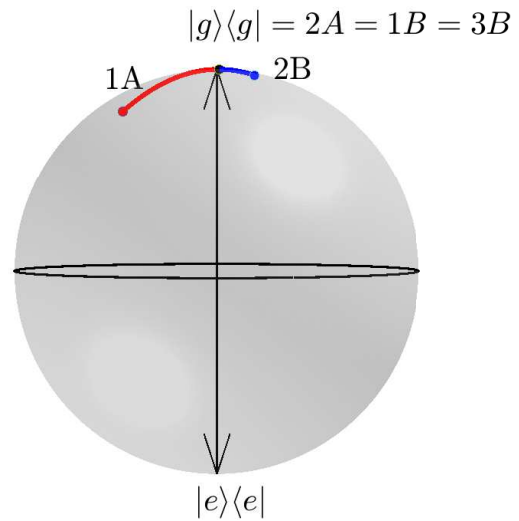}
	
	\caption{Trajectories in the Bloch sphere. 
		The initial reduced state is a product state of local densities 
		defined by the Bloch vectors $\vec{B}_{a}=(0,0.5,0.8)$ 
		and $\vec{B}_{b}=(0,0,1)$, and $g(R)=\gamma/\gamma_{0}=0.8$. 
		The environment temperature is zero.}
	\label{f:f4}
\end{figure}

In Fig. \ref{f:f4} we show the trajectories followed by 
the states of both atoms in the Bloch sphere. 
We observe that as the atom A releases energy, its state evolves 
from its initial state, $1A$ towards the ground state $2A$. 
On the other hand, since atom B starts in the unexcited 
state $1B$, it initially absorbs energy, which drives it out of 
the ground state. 
Upon reaching the point $2B$, the energy emitted equals 
the energy absorbed, and, from that moment on, emission exceeds 
absorption and the atom relaxes to the ground state ($3B$). 
In the following we will interpret these facts, within the framework 
of each paradigm.

\section{Final Remarks and conclusions}
In this work we have explored a new proposal about the nature 
of heat and work in the quantum regime, explicitly developing 
the comparison, for the simple case of a two-level system, 
of the new definitions with one of the more accepted paradigms 
involving these quantities. 
In addition to making possible to reproduce all the classic results for incoherent states, the predictions based on the 
new approach present several advantages with respect to the 
standard paradigm.
In particular, the new concept of heat is closer to the 
classical one, according to which heat is the part of the energy 
exchange that involves a change in the entropy of the system.

In regard to the definition of work, the main difference in 
relation to the previous paradigm is the possibility of 
obtaining work even if the local Hamiltonian of the system 
does not vary in time. 
This is consistent with situations in which the interaction 
with the environment is time-dependent, so work is expected 
to be performed on the system.


In addition, the new paradigm allows to classify the work 
in two contributions: one associated to Hamiltonian driving,
which coincides with the previous definition, and an additional
one, associated to the coherence of the state of the system. 
The latter, that does not appear as work in the first paradigm, 
is related to the work associated to the rotation of 
the spin direction. 
This contribution is analogous to the classical work required
to rotate a magnetic dipole in an external magnetic field. 

Another important advantage of the new definition of work 
is that it highlights the importance of coherence as a resource, 
an aspect already reported in numerous references. 
In particular, the analysis of a pure dephasing process 
from the new perspective allows us to understand in a new way 
why the ability to perform work decreases 
as the system evolves to the passive state.


A remarkable prediction of the new paradigm is that the 
thermalization process can no longer be understood 
as been a purely thermal process, but rather 
involves a mechanical component associated to the change in 
orientation of the Bloch vector, or, in the general case, 
to a rotation of the eigenstates of the system. 
However, we must not forget that the notion of heat arises 
due to the impossibility of accessing the microscopic 
degrees of freedom of macroscopic systems. 
It is intuitive that, in the absence of that limitation, 
many energy exchange processes could be exclusively 
associated to the concept of work.
In the new paradigm, heat is linked to a more fundamental 
inaccessibility, which is a consequence of the quantum 
description of open systems in terms of mixed states.


Finally, a counterintuitive aspect of paradigm 1 that is not 
resolved by the new proposal has to do with the fact that, since 
they are defined using local variables, heat and work cannot be 
considered energy flows in a strict sense. 
In the general case, the heat released by one part of the system 
can be different from the heat absorbed by the other, and the same 
occurs with the work. 
Maybe these concepts can be defined unambiguously, and respecting 
all the intuitive requirements, only in particular situations, such as those
in which the system of interest interacts with systems which, 
by construction, can only exchange either heat or work in the 
classical sense.

\section*{Acknowledgments}
This work was partially supported by Comisi\'on Acad\'emica de 
Posgrado (CAP), Agencia Nacional de Investigaci\'on e Innovaci\'on  (ANII) and Programa de Desarrollo de las Ciencias B\'asicas (PEDECIBA)  (Uruguay). The authors thank Borhan Ahmadi for stimulating discussions.


\begin{thebibliography}{99}
	\bibitem{Carnot} S. Carnot, Reflections on the Motive Power 
	of Fire and on Machines Fitted to Develop that Power, 
	Paris: Bachelier (1824).
	
	\bibitem{Kosloff} R. Kosloff, 
	Entropy \textbf{15}, 2100 (2013).
	
	\bibitem{Sonntag} R.E. Sonntag, C. Borgnakke and G.J. Van Wylen,
	 \emph{Fundamentals of Thermodynamics} (John Wiley, New York, 2003).    

	\bibitem{Zemansky} M.W. Zemansky and R.H. Dittman, \emph{Heat and Thermodynamics}
	 (McGraw-Hill, Auckland, 1997).

	\bibitem{Callen} H.B. Callen, \emph{Thermodynamics and an Introduction to
	 Thermostatistics} (John Wiley, New York, 1985).
	 
	\bibitem{Cengel} Y.A. Cengel, and M.A. Boles, 
	\emph{Thermodynamics: An Engineering Approach}, 4th edition,
	1056 pp. ( McGraw-Hill, New York, 2001).
	
	\bibitem{Bender} C. M. Bender, D. C. Brody, and B. K. Meister,
	 J. Phys. A: Math. Gen. 33, 4427 (2000).
	 
	\bibitem{Jarzynski} C. Jarzynski, J. Stat. Mech.: Theo. Exp., 2004, P09005
	(2004).

	\bibitem{Uzdin} R. Uzdin, A. Levy, and R. Kosloff,
    Phys. Rev. X \textbf{5}, 031044 (2015).

	\bibitem{Bera} M. N. Bera, A. Riera, M. Lewenstein, and A. Winter,
	Nature Commun. \textbf{8}, 2180 (2017).
	
%
	
	\bibitem{Horodecki} M Horodecki, J Oppenheim, Nat. Commun. \textbf{4}, 2059 (2013).
	
	\bibitem{Skrzypczyk} P. Skrzypczyk, A. Short and  J. Popescu, Nat. Commun. \textbf{5}, 4185 (2014).
	
	\bibitem{Alicki} R. Alicki, J. Phys. A: Math. Gen. \textbf{12}, L103 (1979). 
	
	\bibitem{Deffner} S. Deffner and E. Lutz,
	 Phys. Rev. Lett. \textbf{107}, 140404 (2011).
	 
	\bibitem{Alicki2} R. Alicki, M. Horodecki, P. Horodecki, and R. Horodecki,
	  Open Syst. Inf. Dyn. \textbf{11}, 205 (2004).
	  
	\bibitem{Ahmadi} B. Ahmadi, S. Salimi, and A. S. Khorashadl, arXiv:1912.01983 [quant-ph].
	  
	\bibitem{Alipour1} S. Alipour, A. Chenu, A. T. Rezakhani, and A. del Campo, 
	arXiv:1912.01939 [quant-ph].
	
	\bibitem{Nielsen} M.A. Nielsen and I.L. Chuang, 
	\emph{Quantum Computation and Quantum Information} 
	(Cambridge University Press, Cambridge, 2000).
	
	\bibitem{Vallejo1} A. Vallejo, A. Romanelli and R. Donangelo,
	 Phys. Rev. E \textbf{101}, 042132 (2020).
	 
	\bibitem{Su}S. Su, J. Chen, Y. Ma, J. Chen, and C. Sun,
	 Chin. Phys. B \textbf{27}, 060502 (2018).
	 
	\bibitem{Francica} G. Francica, J. Goold, and F. Plastina,
	Phys. Rev. \textbf{E} 99, 042105 (2019).
	 
	\bibitem{Baumgratz} T. Baumgratz, M. Cramer, and M. B. Plenio,
	Phys. Rev. Lett. \textbf{113}, 140401 (2014).
	 
	\bibitem{Scully} M. O. Scully, M. S. Zubairy, G. S. Agarwal,
	and H. Walther, Science \textbf{299}, 862 (2003).
	
	\bibitem{Korzekwa} K. Korzekwa, M. Lostaglio, J. Oppenheim, and D. Jennings,
	New J. Phys. \textbf{18} (2), 023045 (2016).
	
	\bibitem{Abiuso} P. Abiuso, H. J. D. Miller, M. Perarnau-Llobet, and
	M. Scandi, Entropy 22, 1076 (2020).
	
	\bibitem{Szilard} L. Szilard, L. D. Über, 
	Zeitschrift für Physik \textbf{53}, 840–856 (1929).
	
	\bibitem{Toyabe} S. Toyabe, T. Sagawa, M. Ueda, E. Muneyuki, and M. Sano,
	Nature Phys. \textbf{6}:988 (2010).
	
	\bibitem{Peterson} J. P. S. Peterson, R. S. Sarthour, A. M. Souza, I. S. Oliveira,
	 J. Goold, K. Modi, 	D. O. SoaresPinto, and L. C. C´eleri, Proc. Royal Soc. A, 
	 \textbf{472}(2188) {2016}.
	
	\bibitem{Gemmer} J. Gemmer, M. Michel, and G. Mahler, 
	Quantum Thermodynamics: The Emergence of Thermodynamic 
	Behavior within Composite Quantum Systems,
    Lecture Notes in Physics, Springer (2004).
	
	\bibitem{Johal} R. S. Johal, Phys. Rev. E, \textbf{80}, 041119 (2009).
	
	\bibitem{Poliblanc} D. Poilblanc, Phys. Rev. Lett. \textbf{105}, 077202 (2010).
	
	\bibitem{Williams}N. S.  Williams, K. Le Hur, and A. N. Jordan,
	J. Phys. A: Math. Theor., 44385003 (2011).
	
	\bibitem{Latune} C. L. Latune, I. Sinayskiy and F. Petruccione, 
	Quantum Science and Technology \textbf{4} (2), 025005 (2019).
	
	\bibitem{Brunner2} N. Brunner, N. Linden, S. Popescu, and P. Skrzypczyk,
	Phys. Rev. E \textbf{85}, 051117 (2012).
		
	\bibitem{Manatuly} A. Manatuly, W. Niedenzu, R. Rom´an-Ancheyta, B. Cakmak, O. E. 							Mustecaplio˘glu, and G. Kurizki, Phys. Rev. E 99, 042145 (2019).
    
    \bibitem{Roman} R. Rom\'an-Ancheyta, B. Çakmak, and Ö. E. Müstecaplioğlu, Quantum Science and 				Technology \textbf{5}, 015003 (2019).
    
    \bibitem{Ali} M. M. Ali and W. M. Zhang, W. M Zhang, Sci. Rep. \textbf{10}, 13500 (2020).
    
    \bibitem{Partovi} M. H. Partovi, Phys. Rev. E \textbf{77}, 021110 (2008).
    
    \bibitem{Micadei} K. Micadei, J. P. S. Peterson, A. M. Souza, R. S. Sarthour, I. S.
	Oliveira, G. T. Landi, T. B. Batalhao, R. M. Serra, and E. Lutz,
	Nat. Commun. \textbf{10}, 2456 (2019).
	
	\bibitem{Salinas} S. R. A. Salinas, \emph{Introduction to Statistical Physics}
	(Springer, New York, 2001).
	
	\bibitem{Pathria} R.K. Pathria, \emph{Statistical mechanics} (Oxford University press, Oxford, 1996).
	
	\bibitem{Mahdavifar} S. Mahdavifar and A. Akbari, 
	J. Phys. Soc. Japan \textbf{77}, 024710 (2008).
	
	\bibitem{Vallejo2} A. Vallejo, A. Romanelli, R. Donangelo, \textit{incoming}.
		
	\bibitem{Breuer} H.-P. Breuer and F. Petruccione, 
	\emph{The Theory of Open Quantum Systems}  
	(Oxford University Press, Oxford, 2002).
	
	\bibitem{Esposito} M. Esposito, K Lindenberg, C. Van den Broeck, 
	New J. Phys., \textbf{12}, 013013 (2010).
	
    \bibitem{Brunelli} 
    M. Brunelli, L. Fusco, R. Landig, W. Wieczorek, J. Hoelscher-Obermaier, G. Landi, 
    F. L. Semiao, A. Ferraro, N. Kiesel, T. Donner, G. De Chiara, and M. Paternostro, 
	Phys. Rev. Lett., \textbf{121}, 160604 (2018).
	 
	\bibitem{Jakobczyk} L. Jak\'obczyk, A. Jamr\'oz, 
	Phys. Lett. A \textbf{318}, 318 (2003).
	
	
	
	
	 
\end{thebibliography}
\end{document}